\documentclass[graybox]{svmult}

\usepackage{type1cm}        
\usepackage{makeidx}         
\usepackage{graphicx}        
\usepackage{multicol}        
\usepackage[bottom]{footmisc}
\usepackage{newtxtext}       %
\usepackage{newtxmath}      
\usepackage{multirow}
\usepackage{dblfloatfix}

\makeindex             

\begin{document}

\title*{Designing a Rashomon Machine: Pluri-perspectivism and XAI for Creativity Support}
\author{Marianne Bossema, Rob Saunders, Vlad Glăveanu, Somaya Ben Allouch}
\institute{Marianne Bossema \at Amsterdam University of Applied Sciences, The Netherlands, \email{m.bossema@hva.nl}
\and Rob Saunders \at Leiden University, The Netherlands \email{r.saunders@liacs.leidenuniv.nl}
\and Vlad Glăveanu \at Dublin City University, Ireland \email{vlad.glaveanu@dcu.ie}
\and Somaya Ben Allouch \at University of Amsterdam, The Netherlands \email{s.benallouch@uva.nl}
}
%
%
\maketitle

\abstract*{While intelligent technologies offer unique opportunities for creativity support, there are fundamental challenges in designing human-centered co-creative systems. Explainable AI (XAI) can contribute when shifting its traditional role from justification (explaining decisions) to exploration (explaining possibilities). Contextual understanding is essential for supporting embodied creativity. Generative Artificial Intelligence (AI) models are fundamentally limited, however, by their reliance on disembodied data. We propose Pluri-perspectivism as a framework for XAI, to bridge the epistemological gap between human and machine, and promote creative exploration. It is a pragmatic, action-oriented solution to guide the system, repurposing XAI methods such as the Rashomon Technique. This facilitates exploring a spectrum of creative possibilities, and the exchange of ``perspectives'' between human and machine. Using Pluri-perspectivism as a framework for XAI, we can reintroduce productive friction and support human agency in human-machine creative collaborations.}

\abstract{While intelligent technologies offer unique opportunities for creativity support, there are fundamental challenges in designing human-centered co-creative systems. Explainable AI (XAI) can contribute when shifting its traditional role from justification (explaining decisions) to exploration (explaining possibilities). Contextual understanding is essential for supporting embodied creativity. Generative Artificial Intelligence (AI) models are fundamentally limited, however, by their reliance on disembodied data. We propose Pluri-perspectivism as a framework for XAI, to bridge the epistemological gap between human and machine, and promote creative exploration. It is a pragmatic, action-oriented solution to guide the system, repurposing XAI methods such as the Rashomon Technique. This facilitates exploring a spectrum of creative possibilities, and the exchange of ``perspectives'' between human and machine. Using Pluri-perspectivism as a framework for XAI, we can reintroduce productive friction and support human agency in human-machine creative collaborations.}

\section{Introduction} \label{sec:Intro}

Creative activities are strongly linked to human well-being, especially in old age, as they support health and vitality, strengthen social connections, and offer experiences of engagement and fulfillment~\cite{flood2007creativity, fancourt2019evidence, mcquade2024examining}. Advancements in the field of Artificial Intelligence (AI) offer unique opportunities, see~\cite{karimi2020creative, singh2025systematic, bossema2023human}, but we need to further investigate how (embodied) AI can effectively support everyday life creativity, particularly with non-experts such as older adults~\cite{bossema2023human}. 

Designing human-centered co-creative systems presents fundamental challenges. The efficiency and probabilistic nature of generative AI differs fundamentally from the open-ended, embodied nature of human creativity~\cite{garcia2025paradox}. In addition, current models such as Vision-Language Models (VLMs) rely on ``disembodied'' data, and lack tacit knowledge~\cite{o2024extending}. This reveals a gap between computational understanding and human creative experiences.

To address this gap, we introduced Pluri-perspectivism as a pragmatic, action-oriented framework [5]. The goal is to guide intelligent co-creative systems, using a) A schematic map of the creative context, defining Social, Semiotic, Spatial, Material, and Temporal dimensions; and b) Instructions for the active exchange of viewpoints through Perspective Taking and Perspective Offering. When applied to machines, Perspective Taking does not imply phenomenological access or lived experience, but refers to the operational alignment of explanations with attentional and action-oriented dimensions as selected in the schematic map. In this chapter, we propose Pluri-perspectivism as a framework for XAI in creative domains, answering the following research questions:
\begin{enumerate}
\item{ How can Pluri-perspectivism scaffold XAI for creativity support?} 
\item{What XAI methods could be used and how could they be integrated?} 
\item{What new research opportunities follow from this concept?}
\end{enumerate}

Aligned with the XAIxArts movement~\cite{bryan2023explainable}, we propose to shift creative XAI from justification (explaining decisions) to exploration (explaining possibilities). We contribute with the concept of a ``Rashomon Machine'', repurposing XAI methods. By scaffolding XAI for embodied creative exploration, we can improve human-machine co-creativity and better support human creative experiences. \\

The remainder of this chapter is structured as follows: Section~\ref{sec:Motivation} describes the motivation for our proposition. Section~\ref{sec:Background} provides the theoretical background. We then answer our three research questions: Section \ref{sec:RQ1} proposes Pluri-perspectivism as a framework for XAI (RQ1), Section \ref{sec:RQ2} details the integration of XAI methods (RQ2), and Section~\ref{sec:RQ3} outlines further research opportunities (RQ3). Sections~\ref{sec:4} and~\ref{sec:5} discuss findings and conclude this chapter.

\section{Motivation} \label{sec:Motivation} 
Based on related works and empirical findings, we describe the tension between the requirements and current technological limitations for creativity support, motivating our proposition.

\runinhead{Design Challenges} There are fundamental challenges to the design of intelligent systems for creativity support. First, there is a paradox between AI's generative potential and its nature as a probabilistic imitator~\cite{garcia2025paradox}. Without explicit fine-tuning, models tend to confirm existing inputs~\cite{sarkar2024ai}, rather than promoting meaningful creative exploration. Second, there is a conflict between the efficiency of AI and the nature of human thought. AI's fast and production-oriented content generation contrasts with the slower interplay of intuitive and analytical thinking that humans use to construct meaning~\cite{kahneman2011fast}. 
This can overwhelm users, leading to passive selection and a loss of agency~\cite{inuwa2023algorithmic}. It has been found that non-experts such as older adults are particularly vulnerable to this, often accepting the first output to avoid the cognitive burden of rephrasing prompts~\cite{zamfirescu2023johnny}. Sivertsen et al.~\cite{sivertsen2024machine} warn that explanations can reinforce a loss of human agency by encouraging reliance on machines taking decisions for us. Third, there is a risk of ``de-skilling'' when offloading cognitive processes to AI~\cite{shukla2025skilling, gerlich2025ai}. Moruzzi argues that making an effort renders a creative process valuable to humans, and involves growth and skill-development. It is suggested to reintroduce productive friction into human-AI co-creative processes. In addition, ``slowing down'' AI by integrating it in creative processes is essential to make room for human creativity~\cite{moruzzi2024user}. 

\runinhead{The Context-Sensitivity Problem} To explore the opportunities of intelligent technology for creativity support, we conducted an participatory study in 2024, a course ``Drawing with Robots,'' for older adults~\cite{11217527}. One setup featured a robot arm enhanced with a camera and a VLM. In a dydadic setting, participants could talk to the robot and invite it to collaborate with them in making physical collages. The robot provided descriptions, stories, and suggestions, to offer inspiration and support reflexivity. On request, the system generated inspirational images on, and the robot added drawings to the artwork. Participants appreciated its contributions, however noted a lack of sensitivity to their own artistic intentions within the creative context. This empirical finding revealed a technological limitation, and the need for a structural solution.

\runinhead{Technological Limitations} To align with the embodied, multisensory nature of human creativity, we used a robot to collaborate on tangible works of art. Yet, we encountered that VLMs are trained on `disembodied' visuon-language data, and have limited understanding of embodied creative experiences. Current AI models such as VLMs can learn rules from unstructured data without the need for explicit programming, and discover the latent rules or structures that shape human practice~\cite{o2024extending}. However, unsupervised learning alone does not provide the contextual understanding needed to support human creativity. While VLMs demonstrate reasoning capabilities based on vision-language data, recent studies indicate significant limitations in their understanding of material, temporal, and spatial attributes~\cite{xiong2024large, chen2024spatialvlm}. 

It is essential to acknowledge that human experiences are phenomenological, while VLMs rely on descriptions and mathematics. To bridge this gap, we need to create an interface for contextual grounding and guide generative models for supporting human creativity.

\section{Background} \label{sec:Background}

\subsection{Creative Experience} \label{subsec:1}
Our research focuses on creativity in everyday life, an essential aspect of well-being and vitality~\cite{fancourt2019evidence, flood2007creativity, mcquade2024examining}. We emphasize the creative process rather than artistic production, and view creativity not as an eminent or professional expertise, but as a form of ``little-c'' and ``mini-c'' creativity~\cite{kaufman2009beyond}. These types of creativity can best be understood through the lens of ``creative experiences,'' defined by Glăveanu \& Beghetto~\cite{gluaveanu2021creative} as \textit{``Novel person-world encounters grounded in meaningful actions and interactions, marked by openness, non-linearity, pluri-perspectives, and future orientation''}. Glăveanu argues that looking for, and acting on difference is central to creativity~\cite{gluaveanu2020sociocultural}. In creative experiences, the engagement with diverse viewpoints is fundamental, whereby ``viewpoint'' refers to any attentional focus or action orientation. By approaching the world in different ways, we discover new possibilities. Possibility Thinking~\cite{gluaveanu2024possibility} has been defined as an action-based orientation toward the possible, situated within a social, material, and cultural world. It involves 1) Awareness of the Possible: Being open to discovering what is possible; 2) Excitement about the Possible: Being motivated to exploring new possibilities; 3) Exploration of the Possible: The pursuit of specific possibilities, including experimentation and trial-and-error~\cite{gluaveanu2024possibility}. In human-machine creative dialogues, the exchange of viewpoints must contribute to awareness, excitement, and exploration of possibilities. By designing XAI for this, we can build intelligent systems that support and foster creative experiences.

\subsection{Creative Context} \label{subsec:2}
Building on the definition of creative experience as ``novel person-world encounters,'' we must understand the ecosystem in which these encounters arise. Frameworks like Csikszentmihalyi’s Systems Model of Creativity~\cite{csikszentmihalyi2014systems} and Glăveanu’s Sociocultural Theory of Creativity~\cite{gluaveanu2020sociocultural} reject the idea that creativity occurs inside a single mind. Instead, they argue it emerges through dynamic interactions in a social and material world. Malafouris’ concept of Creative Thinging~\cite{malafouris2014creative} further emphasizes that agents think through making. This implies that creative contexts are fluid and shaped through interaction. As Sawyer notes regarding collaborative settings such as improvisation~\cite{sawyer2000improvisational}, creativity emerges through interactions in real-time, whereby no single collaborator fully controls or oversees the overall state of the creative context. Creative collaboration depends on what De Jaegher and Di Paolo call Participatory Sense-making~\cite{de2007participatory}, the continuous co-creation of meaning through interaction. This connects to post-cognitivist approaches such as 4E cognition~\cite{newen2018oxford}, arguing that cognition is not an internal, computational process in the brain, but co-created through embodied interactions in a material world. In this view, explaining the possible can be a form of active regulation. We therefore propose that in human-machine co-creativity, XAI should enable the exchange of ``enactive explanations''. This presents a unique challenge because humans and machines perceive and interpret their environment differently. When well-designed, XAI can be pivotal in enabling agents with different natures to ground their understanding, and collaboratively navigate the creative context.

\begin{figure}[b] 
\centering
\includegraphics[width=\linewidth]{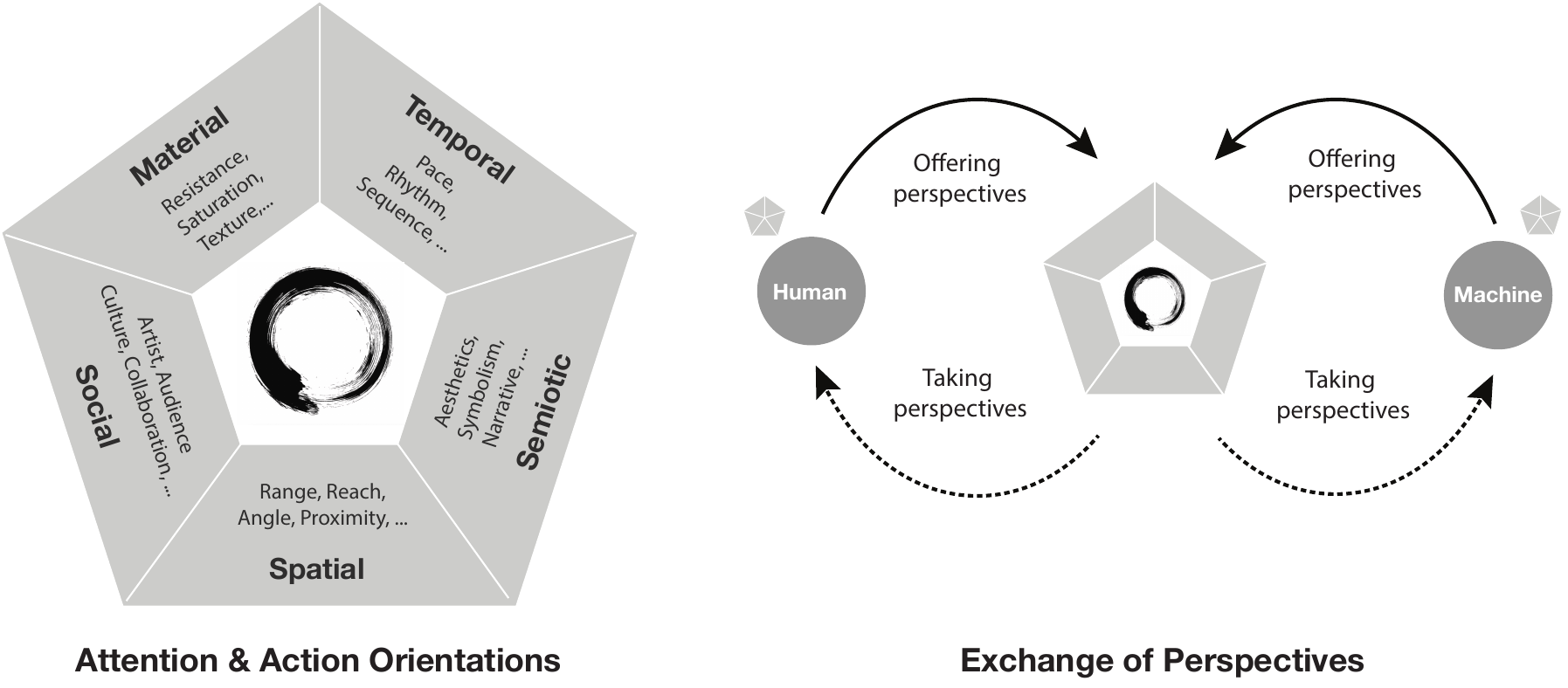} 
\caption{The Pluri-perspectivism Framework for Human-Machine Co-creativity. (Left) Attention \& Action Orientations: A schematic map of the creative context, representing five different orientations towards the artwork in the center. This multidimensional schema serves as a reference system for grounding creative context-sensitivity. (Right) Exchange of Perspectives: The continuous feedback loop of exchanging perspectives, using the schema for mapping orientations and navigating the creative possibility space.}
\label{fig:PAT}
\end{figure}

\subsection{Pluri-perspectivism} \label{subsec:3}
In our empirical research we encounterd the limitation of robot context-sensitivity ~\cite{11217527}. To address this problem, we introduced Pluri-perspectivism as a pragmatic solution to guide a VLM-enhanced robot in a co-creative context~\cite{bossema2026pluri}. The tool builds upon Glăveanu \& Gillespie’s creatogenetic model~\cite{gluaveanu2021creative}, which posits that creativity originates from the tension between three fundamental differences: the Social (Self vs. Other), the Semiotic (Sign vs. Object), and the Temporal (Past vs. Future). The Perspective-Affordance Theory of creativity~\cite{gluaveanu2020sociocultural} connects to this, describing the creative process as a cycle of becoming aware of differences, exchanging positions, and discovering affordances. It defines ``perspective'' as an action orientation that reveals or obscures creative possibilities. Based on this, and building upon theories of 4E cognition~\cite{newen2018oxford} and Participatory Sense-making~\cite{de2007participatory}, we defined 5 dimensional orientations, specifying the embodied creative context: Social, Semiotic, Material, Temporal and Spatial (Figure~\ref{fig:PAT}). The five dimension form a schematic map for exploring a co-creative context. Doyle describes shifts between different creative orientations, affecting one another in a ``figure-ground'' relationship~\cite{doyle2024creative}. For instance, a material affordance may stay in the background until attention shifts, bringing it to the foreground. In creative dialogues, these shifts can be a driving mechanism. In our study into Pluri-perspectivism~\cite{bossema2026pluri}, artists reported to actively use shifts for exploring differences. They experimented with working fast or slow (Temporal), taking more or less space, changing spatial viewpoints (Spatial), or using a variety of materials (Material) to discover new meaning and aesthetics (Semiotics). Similarly, art teachers promoted shifts in orientation with their students, e.g., introducing material unpredictability (Material), or celebrating the diversity of creative expressions within groups (Social)~\cite{bossema2026pluri}. Crucially, the five dimensions are not exclusive categories, but conditioning variables prompting the model to generate a spectrum of explanations. Thereby explanations may bridge multiple dimensions simultaneously. For instance, an artist might report that rotating the canvas (Spatial) revealed a hidden figure (Semiotic), or that the resistance of the paper (Material) forced a slower rhythm (Temporal). By conditioning XAI using this schema, we can guide it in supporting creative shifts. 

\section{Pluri-perspectivism Scaffolding Creative XAI} \label{sec:RQ1}

To answer the first research question, we investigate how Pluri-perspectivism can scaffold XAI, to improve contextual grounding. We build upon In-Context XAI, a method that tailors explanations to specific situations, users, and domain knowledge~\cite{arrieta2020explainable, rong2023towards, liao2021human}. Mill et al.'s SAGE framework~\cite{mill2024sage} addresses the need for defining the XAI context, and provides a structure for a general description. While general frameworks exist, creative XAI asks for on a domain-specific solution. 

\runinhead{Contextual Grounding} XAI can facilitate contextual grounding through the exchange of human and machine explanations, but needs guidance. Our framework structures explanations across five key dimensions: Social, Semiotic, Material, Temporal, and Spatial. By using the dimensions as a schematic map, explanations can be better understood within the embodied creative experience. In addition, the framework facilitates a continuous feedback loop through a) Perspective Taking: Mapping human explanations to interpret human orientations; and b) Perspective Offering: Generating explanations that introduce alternative orientations.

\runinhead{Navigating the Creative Context}
Research showed that navigating an AI model's latent space can support creative exploration, enabling users to traverse dimensions. Unlike domain-specific models, general VLM latent spaces are not explicitly regularized for meaningful attributes. VLMs require extra guidance for creative context-sensitivity. Structured by Pluri-perspectivsm, XAI can offer explanations more strategically, and use human explanations for dynamic refinement. The contextual map serves as an internal model, contributing to machine context-sensitivity. Rather than optimizing for human preferences and risking confirmation bias, the system's explanations can encourage (fluid) shifts in attention and action orientation. This creates productive friction while taking the human perspective as a starting point. Although randomness can yield surprises and new discoveries, Boden and Simonton emphasize that novelty must be assessable and guided by cognitive control~\cite{simonton2018defining, boden2004creative}. Human creativity requires a balance: enough novelty to be visible, and enough familiarity to be recognized as valid~\cite{harvey2023toward}. We argue that when guided by Pluri-perspectivism, XAI can help maintain this critical balance through the exchange of perspectives, facilitating exploration and serendipity.

\runinhead{Technical Implementation}
As described in our Pluri-perspectivsm study~\cite{bossema2026pluri}, we can implement the framework using schema-guided prompting~\cite{lee2021dialogue,zhang2023sgp}. The proposed five-dimensional model then functions as an internal schema for XAI, to map human explanations and generate new ones. Collected human explanations, mapped on the schema, can condition the VLM through few-shot prompting~\cite{brown2020language}. This is a form of in-context learning~\cite{dong2022survey}, allowing the VLM to make inferences based on a schema without requiring retraining. The implementation offers a domain-specific solution, transforming disembodied predictions into contributions grounded in human creative experience.

\section{Integrating XAI Methods for Creativity Support}\label{sec:RQ2}
Turning to the question of method integration (RQ2), we argue that existing XAI techniques can effectively serve as instruments of pluri-perspectival inquiry.

\begin{figure}[t] 
\centering
\includegraphics[width=\linewidth]{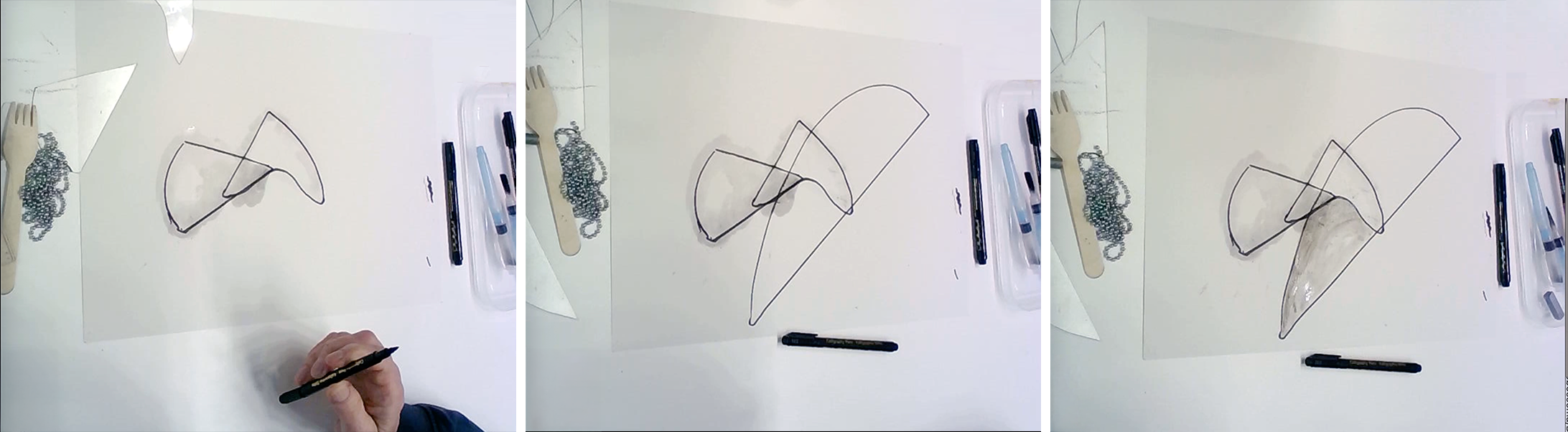} 
\caption{A sequence of snapshots of a drawing activity, showing artwork in progress, providing objects as templates for tracing, water-soluble markers, chalk and water brushes.}
\label{fig:Drawing-Sequence}
\end{figure}

\begin{figure} 
\centering
\includegraphics[width=\linewidth]{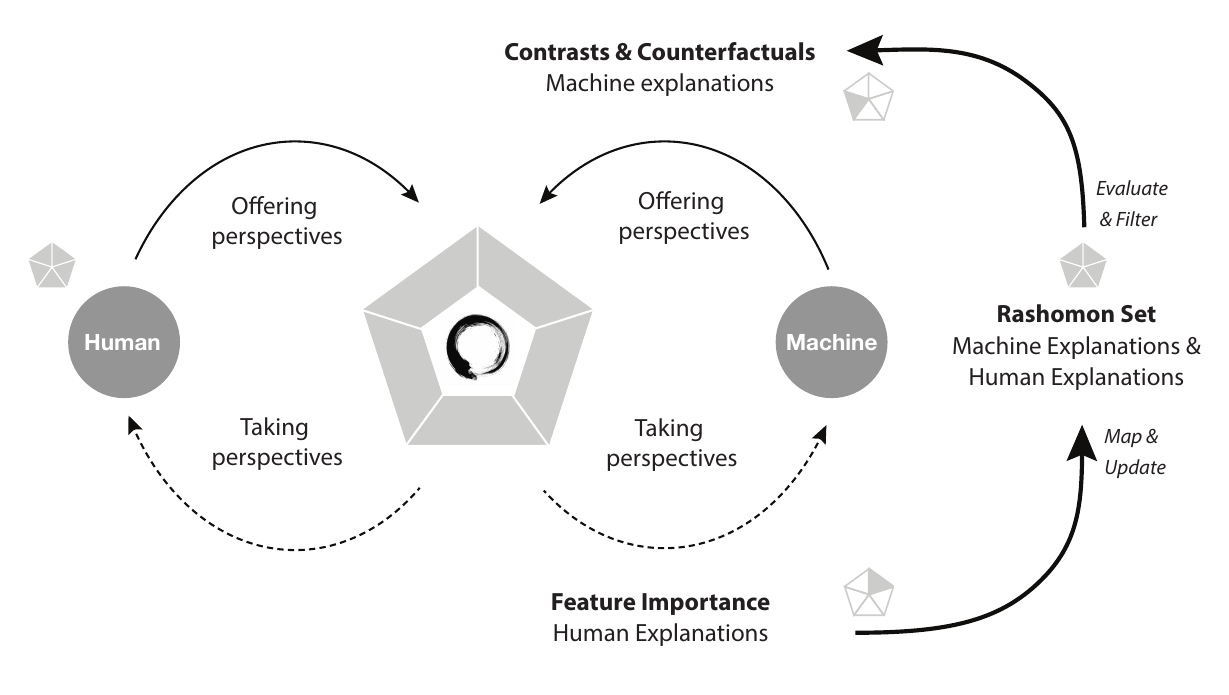} 
\caption{The Rashomon Machine feedback loop with XAI methods facilitating it. Bottom - Perspective Taking: The system assesses Feature Importance and maps user explanations on the Pluri-perspectivism schema. Right: The Rashomon Set functions as a collection of latent possibilities, continuously refined, serving as the system's mental model of the creative context. Top - Perspective Offering: The set is filtered based on the creative state, to select explanations that deepen or broaden exploration.}
\label{fig:XAI-Methods}
\end{figure}

\runinhead{The Rashomon Technique} can be used for generating a set of possible explanations for a given state in the creative process. The Rashomon Effect~\cite{heider1988rashomon} is a concept used in social sciences for situations where multiple contradictory viewpoints and explanations can co-exist. In AI, the Rashomon Technique is a conceptual framework and analysis method, based on the insight that for any given dataset, different models can be equally accurate but solve the problem in different ways. By generating and analyzing a Rashomon Set of all good models, AI researchers can choose the model that best meets their criteria, e.g., of being fair, stable, or interpretable~\cite{breiman2001statistical}. In our case, we produce a set of equally valid but distinct explanations for a single creative state, mapped on the Social, Temporal, Semiotic, Material and Spatial dimensions of our Plur-perspectivism framework. It is important to note that these dimensions serve as generative prompts, not for classification. The dimensions act as lenses to broaden the spectrum of generated explanations, and a generated Rashomon Set can contain explanations that span multiple dimensions. The set acts as the systems mental model of the creative context, to map human explanations, and for selecting alternative explanations promoting exploration. Figure~\ref{fig:Drawing-Sequence} shows a sequence of images captured during a drawing activity, Table~\ref{tab:Initial_set} shows a Rashomon Set based on the current state of this artwork in progress. Each explanation has an attribute, an anchor point that can be used for generating alternative explanations within the same dimension. For instance, an alternative to the first explanation could be ``\textit{Quick and light sketching with chalk will contrast with the deliberate marker lines}''.
\runinhead{Contrastive and Counterfactual Explanations}~\cite{miller2019explanation} allow for Perspective Offering, suggesting divergent trajectories (Figure~\ref{fig:XAI-Methods}). Responses to these explanations can be used to assess the boundaries of human intent and perceived affordances. Contrastive and counterfactual explanations facilitate different strategies. Contrastive explanations answer ''\textit{why this and not that?},'' distinguishing the user's current orientation from alternative options based on the five-dimensional map. Counterfactuals introduce ``what if?'' scenarios, simulating future creative states. Contrastive and counterfactual explanations can trigger shifts between dimensions, or within dimensions using attributes as anchor points. For example: ``\textit{What if you would leave a shape open and I would finish it?}''
\runinhead{Feature Importance} plays a critical role in XAI by identifying which input variables (features) most significantly influence a model's predictions~\cite{lundberg2017unified}. In our case we can use the method for Perspective Taking, to assess the human orientation at a given state. Human expressions reveal an orientation that can be mapped to the Rashomon Set that the system keeps as a mental model. This informs what explanations can be selected or generated for Perspective Offering. The Rashomon Set can be continuously refined and enhanced with human explanations, improving the system's context-sensitivity (Figure~\ref{fig:XAI-Methods}).  

This demonstrates how existing XAI methods, when integrated by the Pluri-perspectivism framework, can serve as the functional instruments for navigating the embodied co-creative context.

\renewcommand{\arraystretch}{1.6}
\begin{table}[!h]
\caption{An initial Rashomon Set of generated explanations, based on Figure~\ref{fig:Drawing-Sequence}. Dimensions are conditioning variables, prompting a spectrum of enactive explanations and affordances. Generated explanations may overlap Dimensions or Attributes, enabling fluid shifts.}
\begin{tabular}{|p{1.4cm}|p{8.1cm}|p{1.7cm}|}
    \hline
    \textbf{Dimension} & \textbf{Enactive Explanation} & \textbf{Attribute}\\
    \hline
    \multirow{3}{*}{Material} & Pressing the marker creates friction to slow the hand. & Resistance \\ \cline{2-3}
            & Adding water dissolves the marker lines into a spreading wash. & Saturation \\ \cline{2-3}
            & Rubbing the chalk side reveals the paper's rough grain. & Texture \\\hline
    \multirow{3}{*}{Spatial} & Extending gestures allows for reaching the paper's edges. & Range\\ \cline{2-3}
            & Shifting the viewpoint changes the perceived overlap of the shapes. & Angle \\ \cline{2-3}
            & Leaning in close directs focus to the intersection of fine lines.& Proximity\\ \hline
    \multirow{3}{*}{Temporal} & Slowing the stroke allows the water-soluble ink to bleed deeper. & Pace\\ \cline{2-3}
            & Dabbing the brush creates a staccato against the continuous lines. & Rhythm\\ \cline{2-3}
            & Drawing the outline before wetting it preserves the structure. & Sequence\\ \hline
    \multirow{3}{*}{Semiotic} & Applying heavy strokes gives the floating forms a sense of gravity. & Aesthetics \\ \cline{2-3}
            & Closing the loop makes the shape represent a protected space.& Symbolism\\ \cline{2-3}
            & The line wanders away to explore the empty white space. & Narrative\\ \hline
    \multirow{3}{*}{Social} & Choosing permanent markers signals a preference for bold contrast. & Artist\\ \cline{2-3}
            & Using fluid washes connects the drawing to watercolor traditions. & Culture\\ \cline{2-3}
            & Leaving the shape open invites the robot to finish it. & Collaboration\\ \hline
\end{tabular}
\label{tab:Initial_set}
\end{table}

\section{Opportunities for Further Research}\label{sec:RQ3}

We address the last question by identifying research opportunities arising from our proposition. We argued that creativity support requires a balance between novelty and familiarity, based on theory~\cite{simonton2018defining, boden2004creative, harvey2023toward}. When guided by Pluri-perspectivism, XAI can maintain this balance through shifts that facilitate exploration and serendipity. But how can the system select explanations adaptively and timely? 

\runinhead{Selection and Evaluation} Our goal is to stimulate Possibility Thinking (awareness, excitement, and exploration of the possible)~\cite{gluaveanu2024possibility}. When a human artist is actively exploring a specific orientation or expressing a motivation to do so, the system could offer contrastive explanations within the same dimension, to deepen exploration. Alternatively, in a case of an impasse, the system could trigger shifts between dimensions to broaden exploration. Future work must explore heuristics for tracking creative states, and define metrics for real-time evaluation. When the system can assess whether explanations supported and enhanced exploration, new explanations can be selected adaptively. 

\runinhead{Taking Initiative} Related to this, there is the challenge of timing. Currently, our framework focuses on what to say, but not when to say it. Defining the dynamics of initiative is critical, and interrupting a state of flow can be undermining. Future XAI systems must model the temporal dynamics of the artistic process to decide when to interrupt. Silence can be a valid XAI action as well, and we need to investigate how we can use it strategically. In human-human collaboration, knowing when not to interrupt is as important as the explanation itself.

\section{Discussion} \label{sec:4}

\runinhead{Theoretical and Methodological Contributions}
We take the stance that to support human creativity, generative AI must be able to align with embodied creative experiences. XAI is pivotal for the collaborative exploration of creative possibilities. In this chapter, we addressed three fundamental questions on how we can design XAI for this. Regarding RQ1, we proposed Pluri-perspectivism as a framework for XAI, to bridge the gap between disembodied generative AI and embodied creative experiences. Addressing RQ2, we conceptualized the Rashomon Machine, showing how XAI methods (specifically Feature Importance, Rashomon Sets, and Counterfactuals) can facilitate the co-creative feedback loop. Finally, concerning RQ3, we identify that future work needs investigation of the creative interaction dynamics. In addition to generating explanations to support exploration, we need to model creative states informing the system on when to interrupt and when to align with or disrupt the human during the co-creative activity.

\runinhead{Limitations and Future Directions} A remaining conceptual tension concerns the metaphorical status of ``perspective'' when applied to artificial systems. While we adopt this language pragmatically, we emphasise that perspectival agency remains asymmetrically distributed between humans and machines. Taking a pragmatic, action-oriented stance, we propose an interface for contextual grounding that guides generative models to better support human creativity. 

While we have presented a conceptual design, the validity remains to be tested in a real context of  human-machine interactions. Empirical studies are needed to validate whether the solution effectively stimulates creativity with non-expert users, such as older adults. To further develop our concept, we must address the challenges identified in section 3.2. To define policies for determining which explanation to offer and when, we are currently conducting an empirical study analyzing human-human creative dyads. By observing how creative coaches time their interventions to support the creative process, we aim to derive computational heuristics for modeling creative states. To evaluate these states, we will develop specific metrics that quantify how creative exploration develops. For example, we could measure the evolution of the Rashomon Set and the degree to which the user adopts generated explanations. In addition, self-reported creativity and engagement can be assessed. This requires further research which we intend to conduct in follow-up studies.

Although we used Gemini 3 for content generation, we recognize the privacy and fine-tuning limitations of closed cloud models. Consequently, our future work prioritizes data sovereignty, and the local deployment of open-source foundation models. With schema-guided instructions, smaller models can perform as well as large commercial models~\cite{feng2023towards}, while local deployment marks a critical step toward ethical, context-sensitive XAIxArts. 

In addition to schema-guided prompting, other Inference-Time Intervention techniques~\cite{li2023inference} could leverage future implementations. Embodied experiences are often implicit and difficult to describe in a prompt. Recent work in Vision Language Action models demonstrates that real-time sensor input can be aligned with internal representations of physical dynamics~\cite{haon2025mechanistic}. Integrating sensor-based activation steering~\cite{sivakumar2025steervlm} offers a promising path for continuously grounding the Rashomon Machine in the spatial and temporal realities of the creative context.

\section{Conclusion}
\label{sec:5}

In this chapter, we argued that XAI is pivotal for human-machine creative exploration. We contribute by introducing Pluri-perspectivism as a framework for XAI, to bridge the gap between limited AI context-sensitivity and embodied creative practice. By mapping the creative context across social, semiotic, material, temporal, and spatial dimensions, Pluri-perspectivism structures XAI for contextual grounding, repurposing existing XAI methods. A Rashomon Set of enactive explanations functions as the system's dynamic mental model of the creative state, defining contextual affordances. From this set, explanations can be selected that align with or diverge from the human perspective. This allows for reintroducing productive friction and preventing the passive consumption often encouraged by generative models. While challenges remain in evaluating creative states, we lay the conceptual foundation for a ``Rashomon Machine'': a system that actively takes and offers diverse viewpoints to promote co-creative exploration in an embodied context. Future work will focus on further investigating this in human-machine interactions.


%
%

\bibliographystyle{splncs04}
\bibliography{articles}

\end{document}